\documentclass[pra,notitlepage,aps,amsmath,floatfix]{revtex4-1}
\usepackage{amsmath,graphicx}
\begin{document}
\def\tr{\rm{Tr}}
\def\la{{\langle}}
\def\ra{{\rangle}}
\def\a{{\alpha}}
\def\e{\epsilon}
\def\q{\quad}
\def\up{\uparrow}
\def\do{\downarrow}
\def\w{\tilde{W}}
\def\t{\tilde{t}}
\def\A{\mathcal{A}}
\def\tA{\tilde{A}}
\def\La{\lambda}
\def\h{\hat{H}}
\def\z{\hat{Z}}
\def\E{\mathcal{E}}
\def\N{\mathcal{N}}
\def\lm{\lambda}
\def\p{\hat{P}}
\def\m{\mu}
\def\ps{\psi_0}
\def\b{\psi_0}
\def\pss{|\psi_0\rightarrow}
\def\R{\text{Re}}
\def\I{\text{Im}}
\def\df{\Delta f}
\def\u{\hat{U}}
\def\l{\ell}
\def\B{\hat{B}}
\def\({\{}
\def\){\}}
\def\C{\hat{C}}
\def\D{\hat{D}}
\def\BB{B_w}
\def\BBB{\tilde{B}_w}
\def\MM{\Up^w}
\def\n{\\ \nonumber}
\def\j{\hat{j}}
\def\alph{a}
\def\Up{\mathcal{M}}
\def\vc{\underline{c}}
\def\vf{\underline{f}}
\title{Inverse quantum measurement problem}
\author {D. Sokolovski$^{a,b}$}
\author {S. Mart\'inez-Garaot$^{a}$}
\author {M. Pons$^{c}$}
\affiliation{$^a$ Departmento de Qu\'imica-F\'isica, Universidad del Pa\' is Vasco, UPV/EHU, Leioa, Spain}
\affiliation{$^b$ IKERBASQUE, Basque Foundation for Science, E-48011 Bilbao, Spain}
\affiliation{$^c$Departmento de F\' isica Aplicada I, Universidad del Pa\' is Vasco, UPV-EHU, Bilbao, Spain}
\date{\today}
\begin{abstract}
Quantum mechanics relates probability of an observable event to 
the absolute square of the corresponding probability amplitude.
It may, therefore, seem that the information about the amplitudes' 
phases must be irretrievably lost in the experimental data. Yet, there are 
experiments which report measurements of wave functions, 
and closely related quantities such as Bohm's velocities and positions of 
bohmian particles. We invert the question, and ask under which 
conditions the values of quantum amplitudes can be recovered from
observed probability distributions and averages.
\end{abstract}
\maketitle
\noindent
{Keywords; {\it quantum particle's  past, transition amplitudes, weak measurements}}
\vspace{0.5cm}
\section{introduction}
Quantum mechanics predicts the probabilities, or frequencies, with which certain observed outcomes, or series of outcomes, 
would occur, should the experiment be repeated under the same conditions.
It does so in a peculiar way. In order to evaluate a probability, first one needs to obtain a complex number, known as the probability amplitude, and take its absolute square \cite{Feynl}. The amplitudes, whether related to wave functions obeying the Schr\"odinger equation, or describing transitions a quantum system makes between available states \cite{Feyn}, are therefore ubiquitous in a quantum mechanical analysis. While the route from amplitudes to probabilities is well known, one can also invert the question, and ask whether 
it is possible to deduce the values of the amplitudes from the measured frequencies, and if so, under which conditions?
There are several reasons why the question, simple as it may seem, deserves further discussion.
\newline
On one hand, the probability amplitudes, whose precise status is still being debated in literature (see, for example 
 \cite{Leif}), are often considered mere computational tools, of no further use, once the desired probability has been calculated.
 This view is supported by observing that, since a probability $P$, and the corresponding amplitude $A$, are related by $P=|A|^2$, 
 information about the phase of the $A$ is irretrievably lost whenever $P$ is measured.
 \newline However, contrary to the above assertion, a recent technique of the so-called ``weak measurements" (WM) (for a review see 
 \cite{WVrev} and Refs. therein) allows one to measure the real and imaginary parts of complex ``weak values" (WV), provided that the measurement is highly inaccurate, or the meter is only weakly coupled to the observed system. It is easy to demonstrate \cite{DS1,DS2}, that the WV, obtained in this manner, can be identified with Feynman's transition amplitudes \cite{Feyn}, or their weighted combinations. Despite earlier claims made, e.g. in \cite{WM1} and \cite{WM2}, WV provide little additional insight into quantum behaviour \cite{DS3}, but the fact that values of certain amplitudes can, after all, be recovered from the experimental data is of some interest. 
\newline
Finally, recent progress in experimental techniques has made possible the use of WM for indirect evaluation of simple wave functions in a chosen representation, or of related quantities, such as Bohm velocities \cite{EXP1}-\cite{EXP4} .
Employing weak measurements may not be the only way to retain the information about the phases, and one might want to look into 
other possibilities as well. The purpose of this paper is to find out which types of amplitudes, for which systems, under which conditions, and by what means, can be, in principle, reconstructed from experimental data. We will also ask what such a reconstruction, once achieved, adds to one's understanding of quantum theory.
\newline 
The rest of the paper is organised as follows. In Section II we revisit the basic rules for constructing probabilities with the help of virtual
(Feynman) paths. In Section III we discuss the difference between the past and present, when dealing with outcomes of consecutive 
measurements. In Section IV the approach is applied to a composite measured system+pointer. In Section V we demonstrate that the values of path amplitudes can be recovered from the statistics of the pointer's readings. Section VI gives a simple two-state example. In Section VII we use the path amplitudes in order to evaluate the initial state of the system.
Section VIII discusses the high accuracy limit, in which the approach of Sections V-VII fails.
In Section IX we analyse the low accuracy limit, and briefly discuss weak measurements and weak values.
In Section X it is shown that it is not possible to recover path amplitudes for a ``one step" history, involving only two measurements.
Section XI discusses the distinctions and similarities between amplitudes, probabilities, and averages. 
In Section XII we stress that by measuring the amplitudes, we gain little further insight into quantum behaviour, 
beyond what has already been said in Section II. Section XIII contains our conclusions.

\section{From amplitudes to probabilities}
We start with the basic postulates of quantum mechanics. Suppose 
we want to know the values, at times $t=t_\l>0$, $\l=1,2,...,L$  of $L$ quantities, represented by operators $\B^\l$,
 acting in an $N$-dimensional Hilbert space. 
In each run of 
the experiment, accurate measurements at $t=t_\l$, will yield a sequence $B^{L}_{i_L} \gets...\gets B^{2}_{i_2}\gets B^{1}_{i_1}$, 
where $B^\l _{i_\l}$, are the eigenvalues of $\B^\l$,
some of which can, in principle, be degenerate. 
It is impossible to foresee the outcome of a particular run, but quantum theory 
is able to predict the probability, $P(B^{L}_{i_L} \gets...\gets B^{2}_{i_2}\gets B^{1}_{i_1})$, with which a particular sequence (path)
will appear, i.e., the frequency with which it will occur if the experiment is repeated many times. 
The recipe the theory offers is a peculiar one: 

(i) First, one needs to find the eigenstates of the operators, 
$\B^\l$, $|b^\l_{i_\l}\ra$, and ensure that all eigenvalues  of $\B^1$ are distinct. 
If so, the first measurement, yielding $ B^{1}_{i_1}$ prepares (pre-selects) the system in a state
$|b^1_{i_1} \ra$, and allows us to proceed with the construction of a statistical ensemble, describing the remaining $L-1$ 
measurements.

(ii) Then a complex valued probability amplitude $A(b^{L}_{i_L} \gets...\gets b^{2}_{i_2}\gets b^{1}_{i_1})$
for the system to ``pass through the states $|b^\l_{i_\l}\ra$"
 is defined as a product
\begin{eqnarray}\label{1}
A(b^{L}_{i_L} \gets...\gets b^{2}_{i_2}\gets b^{1}_{i_1})\equiv A(b^{L}_{i_L} \gets b^{L-1}_{i_{L-1}})\times
... \times A(b^3_{i_3}\gets b^2_{i_2})A(b^2_{i_2}\gets b^1_{i_1})
\end{eqnarray}
where
\begin{eqnarray}\label{1a}
A(b^{L}_{i_L} \gets b^{L-1}_{i_{L-1}})
\equiv \la b^{\l}_{i_{\l}} |\u(t_{\l},t_{\l-1})|b^{\l-1}_{i_{\l-1}}\ra 
\end{eqnarray}
is the amplitude for the system to make a transition from $|b^{\l-1}_{i_{\l-1}}\ra$   to $|b^{\l}_{i_{\l}}\ra$,
between $t=t_{\l-1}$ and  $t=t_{\l}$,
and $\u(t',t)\equiv\exp[-i\h(t'-t)]$ is its evolution operator.

(iii) {\it Born rule.} With all eigenvalues of each operator distinct, the probability of observing 
a sequence $B^{L}_{i_L} \gets...\gets B^{2}_{i_2}\gets B^{1}_{i_1}$   is the absolute square of the amplitude (\ref{1}),
\begin{eqnarray}\label{2}
P(B^{L}_{i_L} \gets...\gets B^{2}_{i_2}\gets B^{1}_{i_1})=|A(b^{L}_{i_L} \gets...\gets b^{2}_{i_2}\gets b^{1}_{i_1})|^2.
\end{eqnarray}
 
(iv){\it Superposition Principle.} 
If, for example, two orthogonal states,  $|b^{n}_{i_n}\ra$ and $|b^{n}_{i'_n}\ra$,
 correspond to the same $B^n_{i_n}$, for some  $1<n<L$, Eq.(\ref{2}) should be modified as
\begin{eqnarray}\label{3}
P(B^{L}_{i_L} \gets...\gets B^{2}_{i_2}\gets B^{1}_{i_1})=
|A(b^{L}_{i_L} \gets...\gets b^{n}_{i_n} \gets... \gets b^{1}_{i_1})+ 
A(b^{L}_{i_L} \gets...\gets b^{n}_{i'_n} \gets... \gets b^{1}_{i_1})|^2.
\end{eqnarray}
In this case, one cannot say through which 
of the two states the system passed at $t_n$.
\newline
The rule is different for the last measurement, $n=L$,
\begin{eqnarray}\label{3}
P(B^{L}_{i_L} \gets...\gets B^{2}_{i_2}\gets B^{1}_{i_1})=
|A(b^{L}_{i_L} \gets...\gets b^{1}_{i_1})|^2+ 
|A(b^{L}_{i'_L} \gets...\gets b^{1}_{i_1})|^2
\end{eqnarray}
since the paths, leading to distinguishable final
outcomes, cannot interfere \cite{Feynl}. 
These rules are readily generalised to the case where several operators have several groups of degenerate 
eigenvalues.

(v) {\it Uncertainty Principle} \cite{Feynl}. If all eigenvalues of an operator $\B^n$ are the same, 
$B^{n}_{i_n} =c$, 
we have
\begin{eqnarray}\label{3}
P(B^{L}_{i_L} \gets...\gets B^{n+1}_{i_{n+1}}\gets  c \gets  B^{n-1}_{i_{n-1}}\gets...\gets B^{1}_{i_1})=
|A(b^{L}_{i_L} \gets...\gets b^{n+1}_{i_{n+1}}\gets  b^{n-1}_{i_{n-1}} \gets...\gets b^{1}_{i_1} )|^2
\end{eqnarray}
and it is impossible to determine the state $|b^{n}_{i_{n}}\ra$ through which the system passes
at $t=t'$, unless all but one amplitudes $A(b^{L}_{i_L} \gets...\gets b^{n}_{i_n}\gets ... \gets b^{1}_{i_1})$ vanish.
\newline
Note that these rules can also be applied to a system prepared in a mixed, rather than pure, initial state
(see Appendix).
\newline
The most peculiar feature is the appearance of a seemingly ``unphysical" complex quantity
$A(b^{L}_{i_L} \gets...\gets b^{2}_{i_2}\gets b^{1}_{i{\tiny }_1})$ which {\it must} be involved,  before the tangible 
frequencies can be accessed. There is still no consensus about the precise status of probability amplitudes (see, for example \cite{Leif}).
(The situation is even more  urgent if the amplitude in question is the wave function, since 
the measurements at $t'$ and $t''$ interrupt its continuous evolution, causing the state to ``collapse".
The many worlds approach \cite{MW} sends the redundant parts of the wave function to 
parallel universes. We, however, will follow \cite{Feynl} in asking only for a recipe to calculate the probabilities.)
\newline
In the following we will be interested in no more than three measurements, $L=2$ and $L=3$.
This will allow us to simplify the rather cumbersome notation in the above equations, and 
 write $\C$, and $\D$
for $\B^{2}$ and $\B^{3}$, $|c_j\ra$ and  $|d_k\ra$ for $|b^{2}_{i_{2}}\ra$ and $|b^{3}_{i_{3}}\ra$, respectively.
We also put $t_1=0$, $t_2=t'$, and $t_3=t''$. There are several possibilities.
For example, it is possible to define a  {\it pre-} and {\it post-selected} sub-ensemble \cite{WVrev}, by running the experiment many times, 
and retaining the statistics only for those cases where the first and the third measurements yield some previously chosen 
$B_i$ and $D_k$. This leaves $C_j$ the the only random variable, whose values occur with the probabilities  
\begin{eqnarray}\label{4}
p[C_j] = P(D_k\gets C_j\gets B_i)/\sum_{j'=1}^NP(D_k\gets C_{j'}\gets  B_i), \q j=1,2,...,N
\end{eqnarray} 
which can be measured directly.
\newline
Our aim is, however, to invert the argument and ask what information about the amplitudes $A(d_k\gets c_j\gets b_i)$
can be gained once the probabilities $p[C_j]$ are measured (see Fig.1). 
With all eigenvalues distinct, Eq (\ref{2}) only allows one to learn something about the modulus 
of $A(d_k\gets c_j\gets b_i)$ (see Fig.2 as an example for $N=3$). To be able to reconstruct the amplitudes with all relevant phases, 
one would need to consider composite systems, comprising both the system and a ``measuring device", and look beyond the accurate ``ideal" measurements. This will be done after a further brief discussion.
\begin{figure}[h]
\includegraphics[angle=0,width=8cm, height= 3cm]{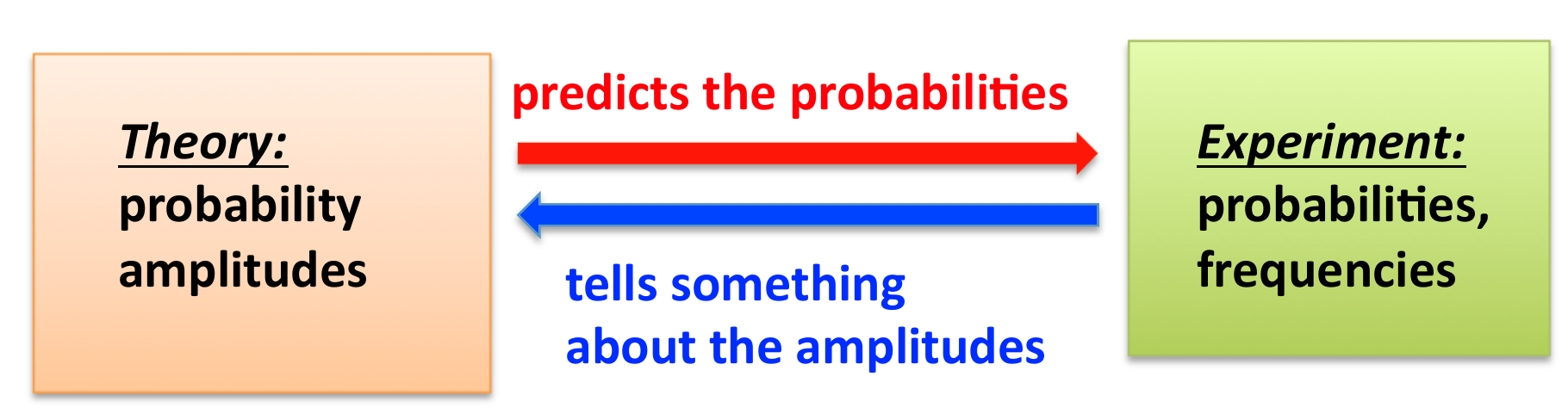}
\caption {A two-way relation between quantum theory an experiment.
Calculated amplitudes predict observable probabilities, while 
measured probabilities allow one to evaluate the amplitudes involved.}
\label{fig:FIG1}
\end{figure}
\begin{figure}[h]
\includegraphics[angle=0,width=8cm, height= 6cm]{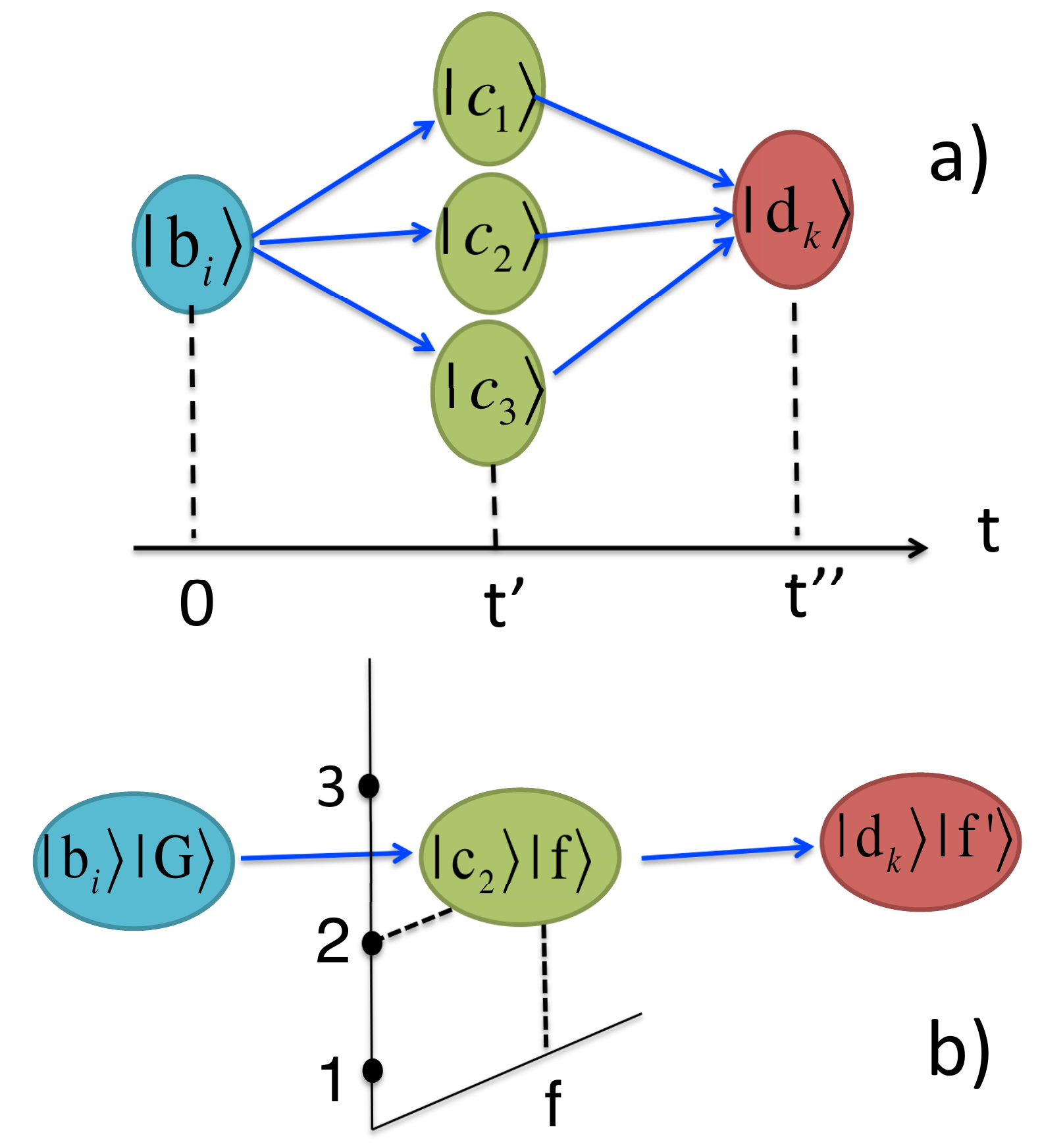}
\caption {a) Three paths, $(d_k\gets c_j\gets b_i)$, $j=1,2,3$, endowed with probability amplitudes
$A(d_k\gets c_j\gets b_i)$ in Eq.(\ref{1}), for $N=3$; b) a path $(d_k,f'\gets c_2,f\gets b_i,G)$ in the Hilbert space of  
the composite system+pointer, furnished with an amplitude $\A(d_k,f'\gets c_2,f \gets b_i, G)$ in Eq.(\ref{c1}), for $N=3$. }
\label{fig:FIG1}
\end{figure}
\section{The difference between the present and the past}
It is worth discussing certain aspects of the above approach, using the three-measurements (two-steps), $L=3$ case as an example.
Firstly, quantum theory provides an amplitude for a path $\{d_k\gets c_j \gets b_i\ra\)$ ``as a whole".
Since, in the sequence,  $c_j$ follows $b_i$ with certainty,
 it does not make sense to ask what is the amplitude for going $c_j\gets b_i$, 
given a later destination $d_k$.  \newline 
 Secondly, as in Section II, quantum theory treats very differently the final moment of a history, 
 a ``now" at $t=t''$, and the ``past", to which the moment $t'$ belongs. To illustrate this, 
 we consider the following simple example.
 Let all three consecutive measurements made on a qubit, $N=2$, be made in the same basis, 
 $|b_i\ra = |d_i\ra =|c_i\ra$, $i=1,2$. With the qubit prepared in a state $|c_1\ra$ by the first measurement at $t=0$, we ask for the values of two quantities whose operators, $\C=\sum_{i=1}^2|i\ra\la i|=\hat 1$ and $\D=
 |1\ra\la1|+|2\ra \la 2|$, obviously, commute. There are four paths, shown in Fig.3, which we will label
\begin{eqnarray} \label{aa3}
\{1,1\} \equiv \{c_1\gets  c_1 \gets c_1 \},\q
\{1,2\} \equiv \{c_1\gets  c_2 \gets c_1 \},\q
\{2,1\} \equiv \{c_2\gets  c_1 \gets c_1 \},\q
\{2,2\} \equiv \{c_2\gets  c_2 \gets c_1 \}.\n
\end{eqnarray}
We note that, since $\C$ is just unity, its value must be $1$ in all cases, with certainty.
Suppose we ask for the values of $\D$, at $t=t'$ and $t=t''$.  By (iii), the probabilities for the four possible outcomes,
$P(I,J)$, $I,J=1,2$ are the absolute squares of the amplitudes $A(I,J)$, ascribed to the paths in Eq.(\ref{aa3}). The probability to obtain $1$ at $t=t''$, regardless of what we have at $t=t'$ is, therefore, 
\begin{eqnarray} \label{aa4}
P(1, \text{anything})=|A(1,1)|^2 +|A(1,2)|^2.
\end{eqnarray}
 Suppose next that we ask about the values of $\C$ at $t=t'$,  and of $\D$, at $t=t''$. By (iii) we have
 \begin{eqnarray} \label{aa5}
P'(1, \text{anything})=|A(1,1) +A(1,2)|^2,
\end{eqnarray}
which is not the same as (\ref{aa4}), since now the paths $\(1,1\)$ and  $\(1,2\)$, corresponding to the same value of $\C$, 
interfere. 
\newline
Finally,  we may ask about the values of $\D$ at $t=t'$,  and of $\C$, at $t=t''$. 
If the paths corresponding to a degenerate eigenvalue of $\C$ at $t=t''$ were to interfere, the probability 
to have both values equal to $1$ would be $P''(\text{anything},1)=|A(1,1)+A(2,1)|^2$. The correct answer, however,  is
 \begin{eqnarray} \label{aa5}
P''(\text{anything}, 1)=|A(1,1)|^2 +|A(2,1)|^2, 
\end{eqnarray}
in accordance with the principle that scenarios leading to distinguishable {\it final} outcomes, never interfere \cite{Feynl}.
The principle is embedded in elementary quantum mechanics as the rule that the mean value of an operator 
$\C$ in a state $|\psi\ra$ is given by $\la \C\ra=\sum_{j} C_j |\la c_j|\psi\ra|^2$, even when the eigenvalues 
$C_j$ are degenerate. With it,
conservation of probability is assured, whichever the order of the measurements. Indeed, since 
$\sum_{j=1}^2|c_j\ra\la c_j|=\hat 1$, $\la c_j|c_{j'}\ra= \delta_{jj'}$, 
\begin{eqnarray} \label{aa6}
P(1, \text{anything})+P(2, \text{anything})=
P'(1, \text{anything})+P'(2, \text{anything})=P''( \text{anything},1)+P''( \text{anything},2)=1,\q
\end{eqnarray}
whereas $|A(1,1)+A(2,1)|^2+|A(1,2)+A(2,2)|^2  \ne 1$.
\newline
The rule that the past outcomes may or may not interfere, depending on the questions asked, 
while the final outcomes are always exclusive \cite{Feyn}, whether we want to know them or not, 
guarantees the consistency of the rules of the previous Section. Suppose we are interested only in the 
values of the first $L-1$ operators $\B^{\l}$, and not in the last $\B^{L}$ at $t=t_L$.
It is sufficient to drop the last term, $A(b^{L}_{i_L} \gets b^{L-1}_{i_{L-1}})$, in Eq.(\ref{1}),
and continue with the remaining amplitudes for the shortened paths $\(b^{L-1}_{i_{L-1}} \gets...\gets b^{1}_{i_1}\)$.
Equivalently, we will get the same answer for the probabilities $P(B^{L-1}_{i_{L-1}} \gets...\gets B^{2}_{i_2}\gets B^{1}_{i_1})$ by simply ignoring the information, obtained at $t=t_L$, 
\begin{eqnarray} \label{aa7}
P(B^{L-1}_{i_{L-1}} \gets...\gets B^{2}_{i_2}\gets B^{1}_{i_1})=\sum_{B^{L}_{i_{L}}}P(B^{L}_{i_{L}} \gets...\gets B^{2}_{i_2}\gets B^{1}_{i_1})\q
\end{eqnarray}
whether or not some, all, or none of the eigenvalues of  $\B^L$ are degenerate. It is, therefore, evident that, 
provided the states $|b^L_{i_L}\ra$ form a complete orthonormal basis, $P(B^{L-1}_{i_{L-1}} \gets...\gets B^{2}_{i_2}\gets B^{1}_{i_1})$ are indeed given by the Born rule (iii), where $t_{L-1}$ becomes the final ``now" time, as described above.
This demonstrates, of course, that the future measurements cannot change the results of the ones already made, thus preserving causality. 
\begin{figure}[h]
\includegraphics[angle=0,height= 7cm]{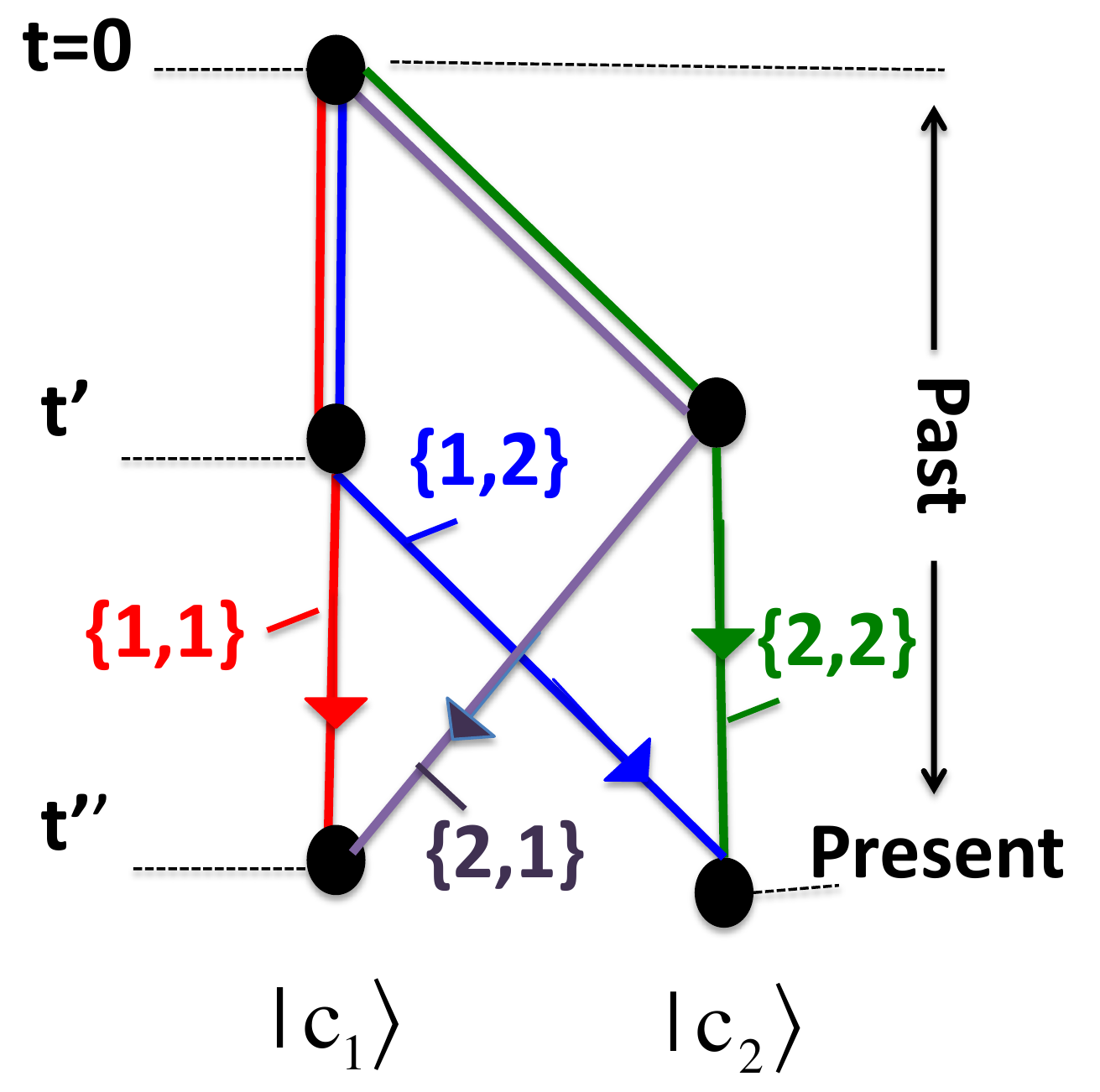}
\caption {Four virtual paths in Eq.(\ref{aa3}). Times $t=0$ and $t=t'$ belong to the ``past", 
while $t=t''$ refers to the ``present", and must be treated differently. }
\label{fig:FIG1}
\end{figure}
\section {Measurements in terms of the amplitudes}
As was mentioned at the end of Section II, in order to recover the amplitudes from the measured probabilities, 
we need to couple the system to another degree of freedom, of a special type. 
One choice of this additional degree of freedom  is a von Neumann pointer \cite{vN}, a one dimensional massive particle with a coordinate $f$, briefly coupled to the system just before $t'$ via $\h_{int}=-i\partial_f \C \delta(t-t')$, where $\delta(x)$ is the Dirac delta. The pointer has no own dynamics apart from $\h_{int}$, so the evolution operator for the composite system+pointer is a product
\begin{eqnarray}\label{c0}
\u(t'',0)=\u_s(t'',t') \exp(-\partial_f \C)\u_s(t',0),
\end{eqnarray} 
where the subscript {\it s} refers to the system. At $t=0$ the system and the pointer are prepared in states $|b_i\ra$ and $|G\ra$, respectively. In what follows, it is convenient, although not necessary,  to think of $G(f)\equiv \la f |G\ra$ as a Gaussian of a width $\Delta f$, 
\begin{eqnarray}\label{c00}
G(f)=(\pi \Delta f^2)^{1/4}\exp(-f^2/2\Delta f^2). 
\end{eqnarray}
\newline
Next we want to describe the work of the measuring device in terms of probability amplitudes, i.e., 
by applying the rules (i)-(v) of Section II to the composite, rather than to the system alone.
First, we define an amplitude for the system and the pointer to pass, just after $t'$, at $t'+0$,  through $|c_j\ra$ and
$|f\ra$, and end up at $t''$ in $|d_k\ra$ and $|f'\ra$, respectively (see Fig. 2b). Using the evolution operator (\ref{c0}), we easily find
\begin{eqnarray}\label{c1}
\A(d_k,f'\gets c_j,f \gets b_i, G)\equiv 
\delta(f'-f) G(f-C_j) A_s(d_k\gets c_j\gets b_i)
\end{eqnarray} 
where
\begin{eqnarray}\label{c2}
 A_s(d_k\gets c_j\gets b_i)\equiv
\la d_k|\u_s(t'',t') |c_j\ra\la c_j|\u_s(t')|b_i\ra,\q
\end{eqnarray}
is the amplitude for the system to follow the path $|d_k\ra \gets |c_j\ra \gets |b_i\ra$ {\it with no pointer present}.
This is what makes the von Neumann pointer special (see also \cite{DSnh}). Dynamical interaction with such a pointer
amounts to destruction of interference between the virtual paths, describing the dynamics of the uncoupled system.
Thus, whatever one may learn about the system, will have to be expressed in terms of the amplitudes (\ref{c2}).
(Note that an interaction of a more general type might bring in terms like $\la d_k|\u_s(t'',t') |c_{j'}\ra\la c_j|\u_s(t')|b_i\ra$, $j\ne j'$, 
not present in the absence of the pointer.) 
\newline
In the end, we will want to look at the final pointer's position at $t=t''$. Its accurate determination may be too difficult, 
so we would need to divide the whole range of $f$ into intervals of a width $\Delta f$, and ask only for the interval, which contains the pointer.
The superposition principle (iv) allows us to construct an amplitude for the pointer to pass through the $m$-th 
interval, rather than through a particular $|f\ra$,  
\begin{eqnarray}\label{c3}
\A(d_k,f'\gets m \gets b_i, G)=
\sum_{j=1}^N\int_{\Delta_m}df  \A(d_k,f'\gets c_j,f \gets b_i, G). 
\end{eqnarray} 
Note that checking whether the pointer passes through $\Delta_m$, amounts to measuring, at $t=t'+0$, the quantity, 
represented by a projector $\hat \pi_m=\int_{\Delta_m} df |f\ra\la f|$, which has eigenstates $|f\ra|c_j\ra$
corresponding to degenerate eigenvalues $0$ and $1$. 
By the Born rule (iii), the probability for the system to end up in $|d_k\ra$, and the pointer 
to pass through $\Delta_m$ at $t=t'+0$, and end up in $|f'\ra$, is given by
\begin{eqnarray}\label{c4}
P(D_k,f'\gets m \gets B_i, G)=|\A(d_k,f'\gets m \gets b_i, G)|^2.\q
\end{eqnarray} 
Finally, since we are not interested in the final pointer's position, the probability to have the system in $|d_k\ra$ at $t''$, 
and the pointer reading at $t'+0$ inside $\Delta_m$, $m\delta f \le f < (m+1)\delta f$, is found by summing 
(\ref{c4}) over all $f'$'s
\begin{eqnarray}\label{c5}
P(D_k\gets m \gets B_i, G)=\int df' |\A(d_k,f'\gets m \gets b_i, G)|^2
=\sum_{j,j'} I_{j,j'} A^*_s(d_k\gets c_{j'}\gets b_i)A_s(d_k\gets c_j\gets b_i), \q\q\q
\end{eqnarray}
where $I_{j,j'}$ is the overlap between the $G$'s centred at $C_j$ and $C_{j'}$, 
\begin{eqnarray}\label{c6}
I_{j,j'}\equiv \int_{\Delta_m} df G(f-C_{j'})G(f-C_{j}).
\end{eqnarray} 
One remark is in order. In deriving Eq.(\ref{c5}) we assumed that the final pointer's position 
$f'$ was measured at $t''$ with the help of the appropriate equipment. However, 
since what we decide to do at a later time cannot affect 
the earlier outcomes, 
we could as well have measured the pointer's final momentum, 
sum over all results, and still obtain Eq.(\ref{c5}). Alternatively, 
at $t=t''$, we could decide to do nothing at all and, as discussed in Section III,  Eq.(\ref{c5})
would still be valid. Thus, Eq.(\ref{c5}) gives a well defined probability 
for the two observable outcomes, $d_k$ and $m$.
\newline
For a measurement to make any sense, it must tell something about the properties of the observed system.
Here these properties are represented by the amplitudes $A_s$'s, describing the system on its own.
What exactly is learnt, depends on the initial pointer's state, and on how the experimental data is processed. 
\section {An inverse measurement problem}
We continue with the pre-Êand post-selected system of Section II, and consider the following experiment. A system,  with an unknown
Hamiltonian, is prepared in the same unspecified state $|b\ra$ and then, at $t=t''$, a detector clicks if it is found in the same unspecified state $|d\ra$ (we dropped the subscripts $i$ and $k$). At $0<t'<t''$ a pointer measures  an operator $\C$, as described in the previous section.
The pointer readings are kept only if the detector clicks  at $t=t''$, and after many trials the experimental data consists
of the numbers of cases, $K(m,d)$,  in which the pointer readings lie in the $m$-th interval, and the system 
ends up in $|d\ra$. These numbers define frequencies, which tend to probabilities, as the number of trials increases, $K(d)=\sum_m K(m,d)\to \infty$, 
\begin{eqnarray}\label{d0}
K(d,n)/K(d) \to p(m,d).  
\end{eqnarray}
 If the pointer position $f$ is determined accurately, $\delta f << \Delta f$, we can introduce the corresponding probability density, which, according to Section IV, has the form
\begin{eqnarray}\label{d1}
  \rho(f) \equiv 
  \lim_{\delta f \to 0}\frac{P(d_k\gets m \gets b_i, G)}{\delta f\sum_m P(d_k\gets m \gets b_i, G) }=
\sum_{i=1}^N G_i^2(f)|\tilde A_i|^2 +2\sum_{j'<j}G_j(f)G_{j'}(f)\R[\tilde A^*_{j'}\tilde A_{j}]
\end{eqnarray}
where
\begin{eqnarray}\label{d2}
\tilde A_j\equiv A^s(d_k\gets c_j\gets b_i)/\N,
\end{eqnarray}
are renormalised system's path amplitudes.
The factor  $\N$ is the probability for the 
system to arrive at (to be found in) $|d_k\ra$ at $t=t''$, regardless of what the pointer reads, 
\begin{eqnarray}\label{d3}
\N \equiv \left \{ \sum_{i=1}^N |A^s_i|^2 +
2\sum_{j'<j} \R[A^{s*}_{j'}A^s_{j}]\int G_j(f)G_{j'}(f)df\right \}^{1/2},
\end{eqnarray}
and $G_j(f)$ is a shorthand for $G(f-C_j)$
\begin{eqnarray}\label{d4}
 G_j(f)\equiv G(f-C_j).
\end{eqnarray}
\newline
The experimentalist wants to obtain information about the system, when it is uncoupled from the pointer.
Clearly, he/she may only learn something about the path amplitudes $A^s(d_k\gets c_j\gets b_i)$, which enter the r.h.s. of Eq.(\ref{d1}).
There are $(N+1)N/2$ unknowns, $\R[\tA^*_{j'}\tA_{j}]$, $j'\le j$, so that measuring $\rho(f_\m)$  at some $f_1\ne f_2\ne...\ne f_{(N+1)N/2}$, 
we obtain a system of linear equations, ($\l=1,2,...,(N+1)/N$)
\begin{eqnarray}\label{d3}
\sum_{j=1}^NG^2_j(f_\m)X_{jj}  +  2\sum_{j'< j}^NG_j(f_\l)G_{j'}(f_\l)X_{jj'}=\rho(f_\m),\q
X_{jj'} \equiv \R[\tA^*_{j'}\tA_{j}].
\end{eqnarray}
Since only the relative phases of $\tA_j$ are of importance, we can choose $\tA$ to be real positive, an then obtain the phases
of all remaining amplitudes, $\tA_j=|\tA_j|\exp(i\phi_j)$,
\begin{eqnarray}\label{d4}
|\tA_j|=\sqrt{X_{jj'}}, \q \phi_j=\cos^{-1}\left (X_{j1}/\sqrt{X_{jj}X_{11}}\right ).
\end{eqnarray}
The result has certain predictive powers. The knowledge of the path amplitudes $A^s(d_k\gets c_j\gets b_i)$ allows one to construct 
a probability density $\rho'(f)$, if a different operator, $\C'$, which commutes with $\C$, $[\C',\C]=0$, is measured for the same
pre- an post-selected system. For a $\C'$, not commuting with $\C$, the scheme would fail due to the appearance
of unknown additional terms $\la d|\u_s(t'',t') |c_{j'}\ra\la c_j|\u_s(t')|b\ra$, $j\ne j'$.
Note also that we have gained no further insight into the origin, nature or usefulness 
of the probability amplitudes. Their role remains as defined by (i)-(v) in Section II, and we are no wiser as to why 
quantum theory must rely on the amplitude in the way it does.
\section{A two-state example}
We illustrate the above on an example which involves a two-level system (qubit).
For $N=2$, at $t=t'$, we can measure $\C=\sigma_z$, where $\sigma_z$ is a Pauli matrix, 
for a particular choice of the initial and final states, $|b\ra$ and $|d\ra$, and an accuracy $\Delta f$.
As it stands, the method of the previous Section is not very practical. 
To evaluate the distribution $\rho(f)$ one would need to count the number of cases, $K(f,f+\delta f)$, 
in which the pointer is found inside $[f,f+\delta f]$, divide it by the total number of trials, $K$, 
and by the interval's width $\delta f$. By the Central Limit Theorem \cite{CLT}, an accurate value
of $\rho(f)$ would be obtained for $K>> 1/\rho(f) \delta f$, i.e., for a number of trials prohibitively large.
It is more convenient to divide the whole range of $f$ into three intervals, $\Delta_I$, $\Delta_{II}$, and $\Delta_{III}$, 
such that the probabilities to find a pointer reading inside each interval are roughly the same, 
$W(Z)=\int_{\Delta_\nu} \rho(f)df \approx 1/3$, $\nu=I,II,III$.
Integrating 
Eq.(\ref{d3}) over $I$, $II$ and $III$,  yields three equations 
for $|\tA_1|$, $|\tA_2|$, and the relative phase $\varphi$,
\begin{eqnarray}\label{e1}
\begin{cases}
J_{11}^{(I)} |\tA_1|^2+J_{22}^{(I)} |\tA_2|^2+2J_{12}^{(I)}\R[\tA^*_{1}\tA_{2}] =W^{(I)}\q\\
J_{11}^{(II)} |\tA_1|^2+J_{22}^{(II)} |\tA_2|^2+2J_{12}^{(II)}\R[\tA^*_{1}\tA_{2}] =W^{(II)}\q\\
J_{11}^{(III)} |\tA_1|^2+J_{22}^{(III)} |\tA_2|^2+2J_{12}^{(III)}\R[\tA^*_{1}\tA_{2}] =W^{(III)}\q\q\\
 \end{cases}
\end{eqnarray}
where
$J_{ij}^{(Z)} \equiv \int_Z G(f-C_i)G(f-C_j)$.
To find the path amplitudes $\tA_i$, one would perform $K>>1$ trials, 
and count the number of times, $K(\nu)$, the reading is found to lie inside an interval $\nu$. 
Replacing the probabilities $W^{(\nu)}$ with the relative frequencies, 
$W^{(\nu)}\approx K(\nu)/K$, and solving (\ref{e1}), will yield the values 
of $\tA_i$, up to an unimportant overall phase. 
\newline
An actual measurement can be simulated by calculating the distribution $\rho(f)$ for a chosen 
system and a pointer, randomly sampling a result $f$ from it, and updating the counts $K(I)$, $K(II)$, 
and $K(III)$, as appropriate.
Results of the simulation are shown in Fig.4, for
 the initial an final states  (given here unnormalised),
\begin{eqnarray}\label{e1b}
|b\ra = (1+8i)|\up_z\ra + (2+3i)|\do_z\ra, \q
|d\ra = (3+4i)|\up_z\ra +(2+7i) |\do_z\ra,
\end{eqnarray}
a measured operator $\C=\sigma_z$, an accuracy $\Delta f/|C_2-C_1|=0.5$, 
an evolution operator
$\u_s(t)=\cos(\omega t) -i\sigma_x \sin(\omega t) $, $\omega t'=\pi/3 $, 
and $\omega t''=\pi/2$.
\begin{figure}[h]
\includegraphics[angle=0, height= 7cm]{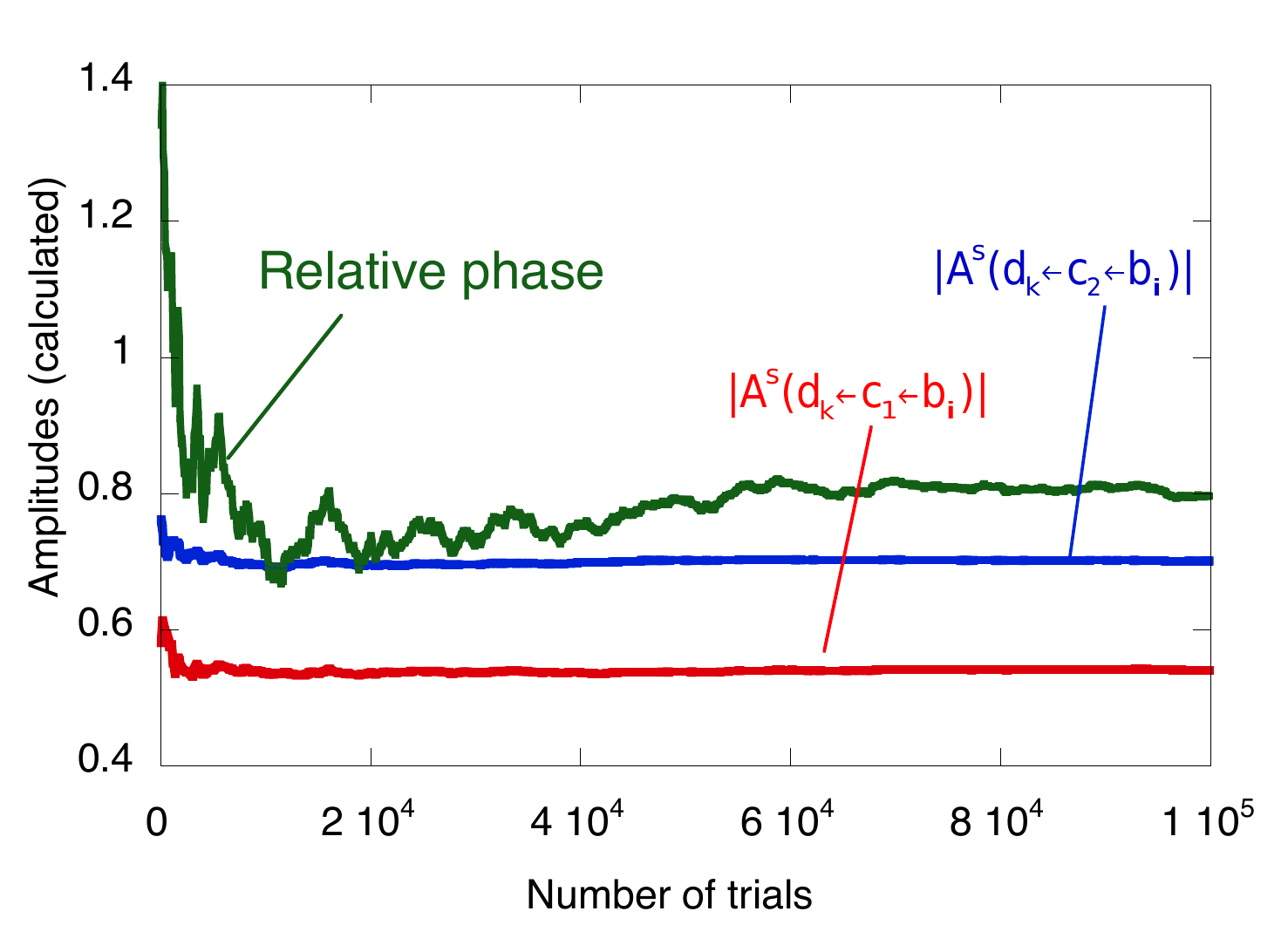}
\caption {Reconstruction of the system's amplitudes from the statistics of the pointer's readings.
After adding $100$ new trials, the values of $|A_1|$, $|A_2|$, and $\varphi$
are recalculated using Eqs.(\ref{e1}). 
The regions $I$, $II$, and $III$, shown in Fig.5, are chosen as
$(-\infty, -0.33 )$, $( -0.33, 0.9  )$ and $(0.9,\infty$), respectively.}
\label{fig:FIG1}
\end{figure}
\newline
Once Eqs.(\ref{e1})  are solved, it is easy to reconstruct the full distribution of the readings, $\rho(f)$, 
or predict this distribution if a different operator, e.g., $\C'$, which commutes with $\C$, is measured to a different accuracy,
 even though nothing is known about the states $|b\ra$ and $|d\ra$ (see Fig.5).
 As discussed in the previous section, it is not possible to predict this distribution for an operator, which does not commute with
 $\C$, because the eigenstate corresponding to an eigenvalue $C_j$,  enters $\tA_j$ only as 
$|c_j\ra\la c_j|$. In particular, for $\C'=\sigma_x$ one would need to know 
 the amplitude for going through $|c'_1\ra=|\up_x\ra=[|\up_z\ra+|\do_z\ra]/\sqrt 2$, 
 \begin{eqnarray}\label{e2a}
A^s(d_k\gets c'_1\gets b_i) =A^s(d_k\gets c_1\gets b_i)/2+
 A^s(d_k\gets c_2\gets b_i)/2+\n
 \la d_k|\u^s(t'',t')|c_1\ra\la c_2| \u^s(t')|b_i\ra/2+
 \la d_k|\u^s(t'',t')|c_2\ra\la c_1| \u^s(t')|b_i\ra/2.
\end{eqnarray}
While solving Eqs.(\ref{e1}) provides the values of the first two terms in the r.h.s. of (\ref{e2a}), up to a phase factor, the last two terms remain unknown.
\section{Indirect ``measurement" of a wave function}
The above approach becomes more useful if the final state $|d(t')\ra=(\u_s)^{-1}|d\ra$ is known, 
\begin{eqnarray}\label{x1}
|d(t')\ra=\sum_{j=1}^N \la c_j|d(t)\ra |c_j\ra,
\end{eqnarray}
while the initial one, $|b\ra$, is not. 
Then obtaining the amplitudes $\tA_j$ by solving Eqs.(\ref{e1}), which are easily generalised to  $N>2$, 
and dividing the results by $\la c_j|d_k(t')\ra$ yields a decomposition of the initial state at the time,
$|b(t')\ra \equiv \u_s(t')|b\ra$,  
of the measurement, 
\begin{eqnarray}\label{x1}
|b(t')\ra=\sum_{j=1}^N\la c_j|b(t')\ra |c_j\ra ,\q
\la c_j|b(t')\ra=\frac{\tA_j}{\la c_j|d(t')\ra}\left [ \sum_{j'=1}^N \left |\frac{\tA_{j'}}{\la c_{j'}|d(t)\ra}\right |^2\right]^{-1/2}.
\end{eqnarray}
Once the unknown state of a system is identified, many other predictions are possible.
Similar, although not identical  techniques were applied to photons, for example in \cite{EXP2} and \cite{EXP3}, 
where the reconstructed wave function was used to evaluate Bohm's velocities, and bohmian trajectories.
\newline
Returning to the Gaussian initial state of the pointer in Eq.(\ref{c00}), we note that $\Delta f$ describes the quantum uncertainty 
in the initial pointer's position. It, therefore, determines the accuracy (resolution) of the measurement even if the pointer's  
final position is determined accurately. 
We note that Eqs.(\ref{e1}) cease to be helpful in the limit of a highly accurate, and therefore, strongly perturbing measurement, 
\begin{eqnarray}\label{e2}
\Delta f \to 0, \q G^2(f-C_j) \to \delta(f-C_j),
\end{eqnarray}
as well as for a highly inaccurate, weakly perturbing one,
\begin{eqnarray}\label{e3}
\Delta f \to \infty, \q G(f-C_j) \to G(f).
\end{eqnarray}
In both limits the matrix, constructed from the $G$'s in the l.h.s. of Eqs.(\ref{e1}), is singular and the method of Sections V and VI fails, as illustrated in Fig.6a. 
\begin{figure}[h]
\includegraphics[angle=0,height= 6cm]{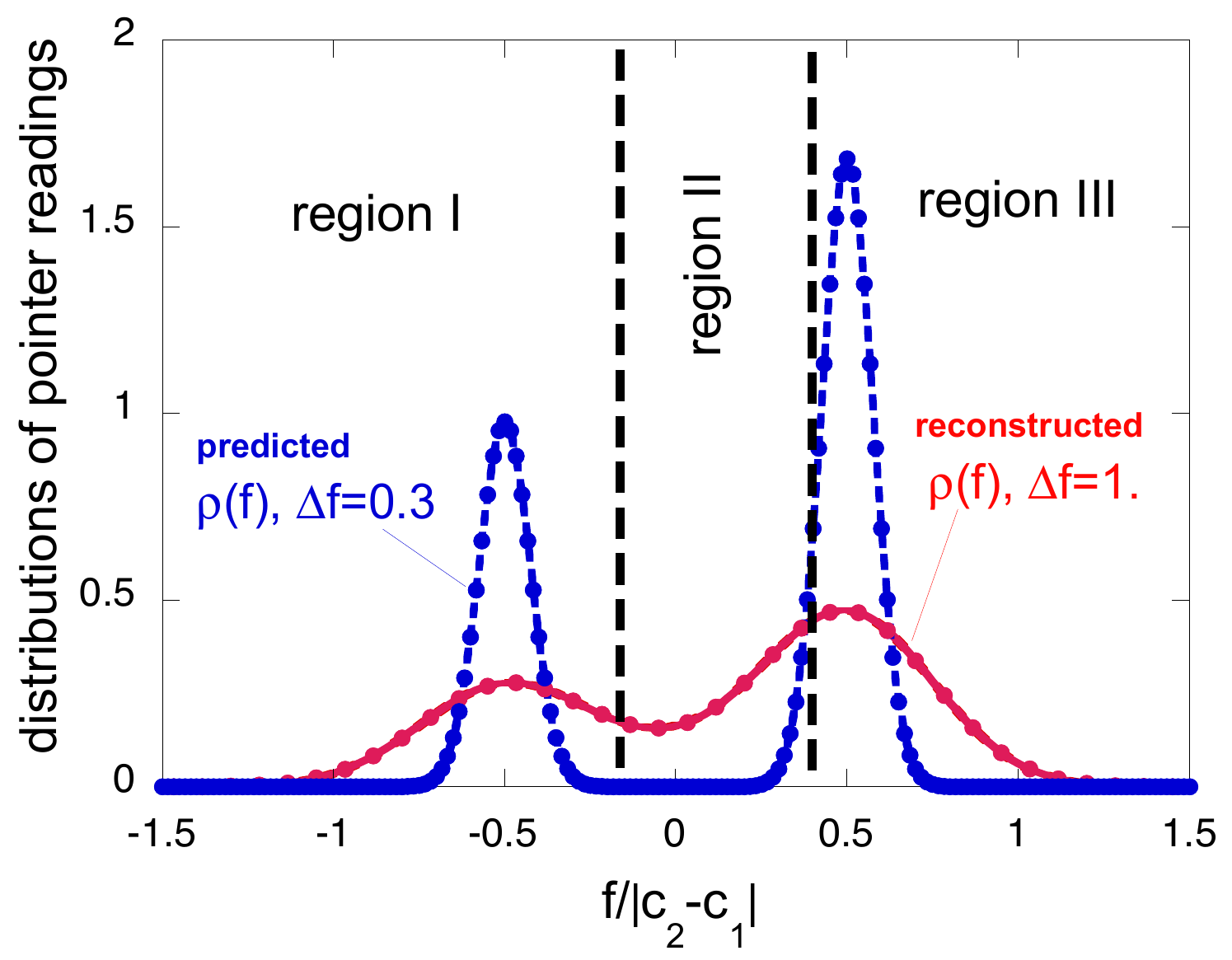}
\caption {Pointer readings distributions $\rho(f)$ for measuring $\sigma_z$ at $t=t'$ for different resolutions, reconstructed from the values used in Fig.3, taken at $K=10^5$ (dots).
The dashed lines are the exact distributions. The three regions $I$, $II$, and $III$, used in the calculation of Section VI, 
are marked by the vertical dashed lines.  }
\label{fig:FIG1}
\end{figure}
\begin{figure}[h]
\includegraphics[angle=0,height= 8cm]{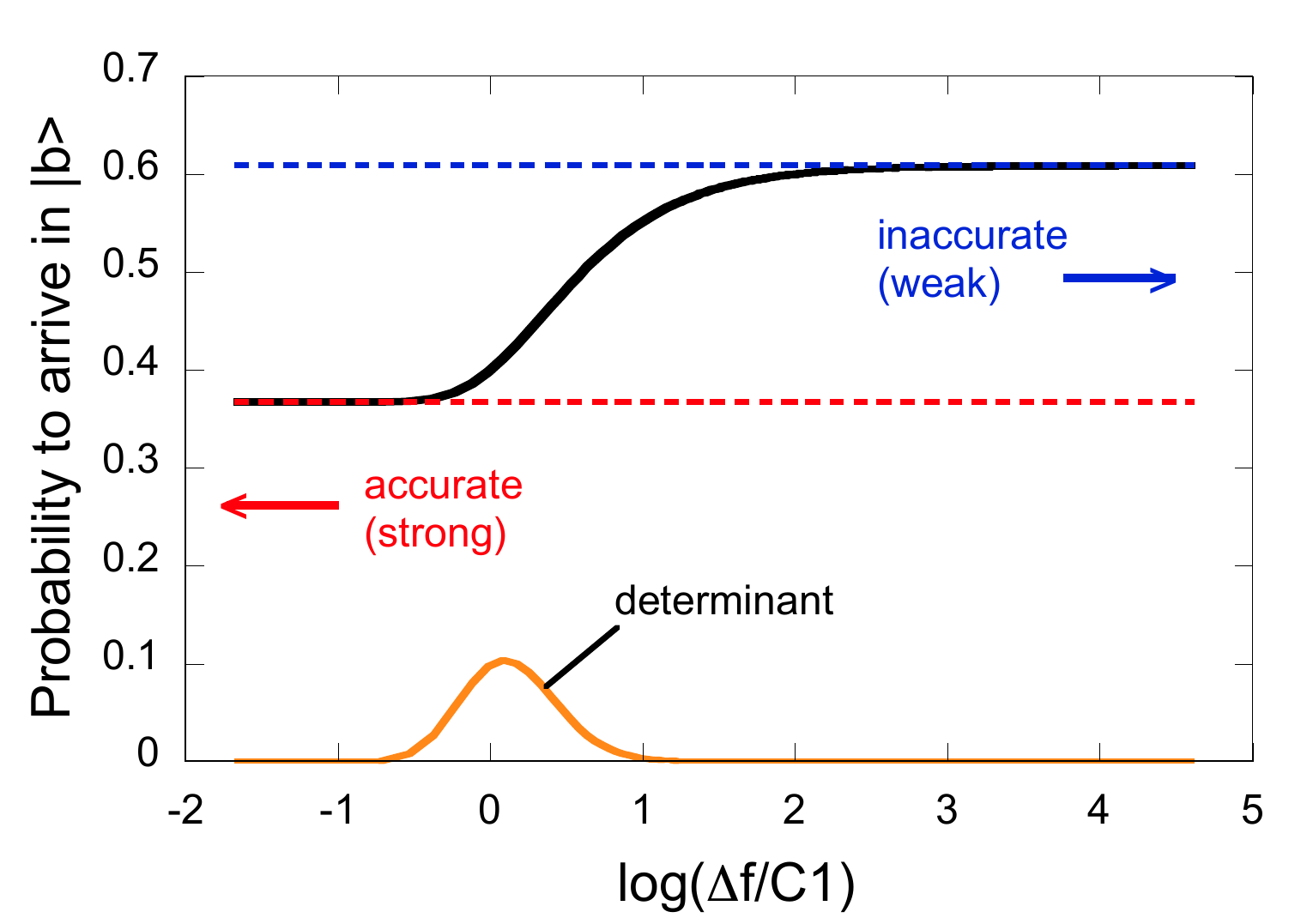}
\caption {a) The results of solving Eqs.(\ref{e1}) numerically for different resolutions $\Delta f$, and 
a fixed number of trials $K=10^5$.
Also shown is the absolute value of the determinant of the matrix (multiplied by 3 for better observation) in the l.h.s. of Eqs.(\ref{e1}).
b) The probability to arrive to the final state $|d\ra$, as a function of $\Delta f$.
The dashed lines mark the limits in which the system is unperturbed by the pointer, 
and in which interference between the paths $(d\gets c_j \gets b)$ is completely destroyed.}
\label{fig:FIG1}
\end{figure}
Next we look at these two limits separately.
\section{Accurate (strong) measurements and the Born rule}
To have an accurate measurement, it is not sufficient to accurately determine the final pointer's position.
One also needs to know its initial setting. The uncertainty of the pointer's initial position is 
proportional to the width of its initial state, $\Delta f$, so that Eq.(\ref{e2}) does define a very precise measurement.
The measurement is also ``strong", since it maximally perturbs the system, whose odds on being detected in the
final state have been significantly altered, as shown in Fig.6b. With all $C_j$ distinct, for the distribution of the pointers's readings, 
 from (\ref{e2}) and (\ref{d1}), we have (restoring subscripts $i$ and $k$)
\begin{eqnarray}\label{f1}
\rho(f)=\frac {\sum_{j=1}^N|A^s(d_k\gets c_j\gets b_i)|^2\delta(f-Cj)}{\sum_{j=1}^N|A^s(d_k\gets c_j\gets b_i)|^2},
\end{eqnarray}
so that the probability to find a reading in a small vicinity of a $C_j$ is $|A^s(d_k\gets c_j\gets b_i)|^2$.
This provides a consistency test for the Born rule, since a measurement always finds the system travelling  
one of the  paths $d_k\gets c_j\gets b_i$,  with a probability proportional to $|A^s(d_k\gets c_j\gets b_i)|^2$.
Counting the number of times, $K(C_j,d_k)$, a result $C_j$ is observed, one constructs the average of $\C$ over many
trials, 
\begin{eqnarray}\label{f2}
\la\C\ra \equiv \frac{\sum_{j=1}^N C_j K(C_j,d_k)}{\sum_{j=1}^N K(C_j,d_k)} \to 
\frac {\sum_{j=1}^NC_j|A^s(d_k\gets c_j\gets b_i)|^2}{\sum_{j=1}^N|A^s(d_k\gets c_j\gets b_i)|^2}.\q
\end{eqnarray}
\newline
In the special case of a projector on a state $|c_n\ra$, $\C=|c_n\ra\la c_n|$, $C_j=\delta_{jn}$, with an $(N-1)$-fold degenerate zero eigenvalue, 
 the average (\ref{f2}) yields the probability to travel the path $d_k\gets c_n\gets b_i$, which is different from 
 what it would be, if all $C_j$ were distinct, 
 \begin{eqnarray}\label{f3}
\la\C\ra=
\frac {|A^s(d_k\gets c_n\gets b_i)|^2}{|A^s(d_k\gets c_n\gets b_i)|^2+|\sum_{j\ne n}A^s(d_k\gets c_j\gets b_i)|^2}
\ne \frac {|A^s(d_k\gets c_n\gets b_i)|^2}{\sum_{j=1}^{N}|A^s(d_k\gets c_j\gets b_i)|^2}.\q\q\q\q\q
\end{eqnarray}
It is readily seen that an attempt to reconstruct the amplitudes from the results of accurate measurements must fail, since 
 $\rho(f)$ in Eq.(\ref{f1}) only contains absolute squares of the $A^s$, and  makes no mention of their phases. (Some phase information
 is retained in Eq.(\ref{f3}), but we will not pursue it further).
\section{Inaccurate (weak) measurements and the relative amplitudes}
Choosing the initial state of the pointer extremely broad, as suggested in Eq.(\ref{e3}), has the obvious inconvenience of spreading
the pointer readings all over the place. It also has an advantage, should  we want it, of perturbing the system only slightly. 
The odds on finding the system in $|d_k\ra$ at $t''$ are almost the same as if the pointer were not there at all (see Fig. 4). 
In the limit $\Delta f \to \infty$ we can expand $G(f-C_j)$ up to the  leading  terms in $\Delta f^{-1}$, $G(f-C_j)\approx G(f)-\partial_f G(f) C_j$, so that the distribution of the readings (\ref{d1}) takes the form
\begin{eqnarray}\label{g1}
\rho(f)= G^2(f)- 2G(f)\partial_f G(f)
\R\left[\frac{\sum_{j=1}^{N} C_j A^s(d_k\gets c_j\gets b_i)}
{\sum_{j=1}^{N}A^s(d_k\gets c_j\gets b_i)}\right]+o(\Delta f^{-2}).
\end{eqnarray}
Although the second term in the r.h.s. is very small ($\sim \Delta f^{-1}$),  it is also very broad ($\sim \Delta f$). Therefore, it
is capable of producing an appreciable change in the averages $\la f^m \ra\equiv\int f^m\rho(f) df$, $m=1,2,..$.
In particular, recalling that  $\int f G^2(f)=0$, for the average pointer reading we find 
\begin{eqnarray}\label{g2}
\la f \ra=\R[\sum C_j\alpha_j] =\R\frac{\la d_k(t')|\C|b_i(t')\ra}{\la d_k(t')|b_i(t')\ra},
\end{eqnarray}
where we defined the relative amplitudes,
\begin{eqnarray}\label{g3}
\alpha_j\equiv \frac{A^s(d_k\gets c_j\gets b_i)}{\sum_{j'}A^s(d_k\gets c_{j'}\gets b_i)},
\end{eqnarray}
and, as before, wrote $|b_i(t')\ra\equiv \u_s(t')|b_i\ra$ and $|d_k(t')\ra\equiv \u_s^{-1}(t'',t')|b_i\ra$.
Historically \cite{AV}, the fraction in the r.h.s. of Eq.(\ref{g3}) was called the WV of the operator $\C$, 
and the reader can follow the developments of the subject in a recent review \cite{WVrev}.
Clearly, the WV is just a sum of path amplitudes, weighted by the eigenvalues of the measured quantity $\C$.
Note that Eq.(\ref{g1}) holds also for the operators whose eigenvalues are degenerate.
\newline
We are still far from our aim to reconstruct the values of the path amplitudes. Firstly, because these occur mixed with
the eigenvalues of $\C$ and, secondly, because only the real part of the sum has been evaluated so far. 
The first problem is easily solved by choosing $\C=|c_n\ra\la c_n|$ to be a projector on $|c_n\ra$. In this case, 
the mean pointer reading coincides with the real part of the corresponding amplitude, 
\begin{eqnarray}\label{g4}
\la f \ra=\R[\alpha_n].
\end{eqnarray}
The second problem is resolved equally easily, if instead of evaluating the pointer's mean position, 
we choose to look at the mean momentum, $\la \lm \ra$, it acquires during the interaction. The result, which can be obtained by 
acting as in Section IV, but considering the paths $(d_k,f'\gets c_2,f \gets b_i)$, is well known \cite{AV,DS1}, and we only quote it here for $\C=|c_n\ra\la c_n|$,
\begin{eqnarray}\label{g5}
\la \lm \ra=2\text{Im}[\alpha_n]\times \int G(f)\partial^2_f G(f) df.
\end{eqnarray}
Using Eqs.(\ref{g4}) and (\ref{g5}), and going over all the projectors $|c_n\ra\la c_n|$ we will be able to evaluate 
all the path amplitudes, normalised to a unit sum, $\sum_{j=1}^N\alpha_n=1$. This is, however, 
a laborious method.
Since the variance of $f$ grows as $\Delta f$ and, by the Central Limit Theorem \cite{CLT},  
the number of trials $K$, required to evaluate  $\R [\alpha_j]$, $\text{Im} [\alpha_j] \sim 1$,  to a sufficient accuracy
would need to grow as $\Delta f ^2$, 
$K>>\Delta f^2$.
\section{One-step histories.  Causality}
Suppose next that we still prepare the system in a state $|b_i\ra$, but are no longer interested in what happens after $t=t'$. There is only one step in the path, $ \(c_j \gets b_i\)$, and the probability to have a value $C_j$ 
of Section II gives 
\begin{eqnarray}\label{h-1}
P(C_j) =|A(c_j \gets b_i)|^2 =\la b_i(t')|\C|b_i(t')\ra,\n
|b_i(t')\ra \equiv \u^s(t')|b_i\ra.
\, \q\q\q\q\q\q\q\q
\end{eqnarray}
Now we apply the same method to a composite system+pointer.
This can happen in at least two ways. Firstly, the machinery designed to determine whether 
the system is in a state $|d_k\ra$ is still in place and is on, but we now use all the results, and do not select 
only those in which the system ends up in $|d_k\ra$. Alternatively, there may be no machine at all, so nothing 
is being done at $t=t''$. In both cases, the distribution of the pointer's readings after $t'$ should be the same, 
as our future decision cannot affect the results already obtained.
\newline
The second option is easily dealt with by recalling that the amplitudes for successive events multiply \cite{Feynl}, 
as is already seen from Eq.(\ref{1}). To obtain the amplitude for the $L-1$ first results, it is sufficient to omit the 
last transitions amplitude, $A(b^{L}_{i_L} \gets b^{L-1}_{i_{L-1}})$ and proceed with the $L-1$ remaining terms as before.
Applying the rules (i)-(v), for the composite system+pointer, we are left with the transition amplitudes
\begin{eqnarray}\label{h1}
\A(c_j,f \gets b_i, G)= \la f|\la c_j|U(t')|b_i\ra|G\ra=
 G(f-C_j)\la c_j|\u^s|b_i\ra, \q\q\q\q\q\q\q\q
\end{eqnarray}
and the observable probability of finding a pointer reading $f$,  
\begin{eqnarray}\label{h2}
\tilde\rho(f)=\sum_{j=1}^N |\A(c_j,f \gets b_i, G)|^2 =
\sum_{j=1}^N G^2(f-C_j)|\la c_j|\u^s|b_i\ra|^2=
\sum_{j=1}^N G^2(f-C_j)|A^s(c_j\gets b_i)|^2.
\end{eqnarray}
Thus, the information about the phases of $A^s(c_j\gets b_i)$ is lost, and cannot be recovered from the distribution 
of the pointer's readings.
\newline
If, on the other hand, there is a machine which detects the system in one of the states $|d_k\ra$, $k=1,2,...,N$,
and the probabilities $P(D_k\gets m \gets B_i, G)$ in Eq.(\ref{c5}) have been measured and recorded, 
the unconditional probability for finding a reading $f$ (we send $\delta_f \to 0$, as in Section IV) is obtained 
by summing over all final states, which yields the same $\tilde\rho(f)$ as in Eq.(\ref{h2}) 
\begin{eqnarray}\label{h3}
\tilde\rho(f)=\sum_mP(d_k\gets m \gets b_i, G)/\delta f=
\sum_{j=1}^N G^2(f-C_j)|A^s(c_j\gets b_i)|^2,   
\end{eqnarray}
since $\sum_{k=1}^N|d_k\ra\la d_k|=1$. This illustrates the causality principle. Two very different
scenarios, played out in the lab a later time, cannot affect the results already obtained.
\newline
Several well known results for measurements  without post-selection follow from Eqs.(\ref{h2}) and (\ref{h3}).
The distribution $\tilde\rho(f)$ can be written in terms of the state of the system immediately before 
it interacts with the pointer, $|\psi(t')\ra\equiv \u^s(t')|b_i\ra$, as
\begin{eqnarray}\label{h4}
\tilde\rho(f)=\sum_{j=1}^N G^2(f-C_j)|\la c_j|\psi(t')\ra|^2
\end{eqnarray}
from which it follows  (since $\C=\sum_{j=1}^N |c_j\ra C_j\la c_j|$ and $\int f G^2(f-C_j)df=C_j$)  that
\begin{eqnarray}\label{h5}
\la f\ra \equiv \int f \tilde\rho(f)df=\la \psi(t')|\C|\psi(t')\ra,
\end{eqnarray}
for a measurement of any accuracy $\Delta f$. 
\newline
A measurement, yielding a result  $f$, leaves the system in a state (we omit normalisation)
\begin{eqnarray}\label{h6}
|\psi(t+0)\ra \sim  \sum_{j=1}^N G(f-C_j)\la c_j|\psi(t')\ra |c_j\ra.
\end{eqnarray}
Thus, a very accurate measurement (Eq.(\ref{e2})) ``projects" the measured system onto one of the states
$|c_j\ra$ with a probability $|\la c_j|\psi(t')\ra|^2$.
A highly inaccurate measurement Eq.(\ref{e3}), yielding a result $f$, perturbs the system only slightly, 
\begin{eqnarray}\label{h7}
|\psi(t+0)\ra \sim [G(f)-\sum_{j=1}^N \partial_fG(f)\C] |\psi(t')\ra +o(\Delta f^{-1}),\q\q
\end{eqnarray}
and the mean value of an operator $\C$ in Eq.(\ref{h5}), $\la \psi(t')|\C|\psi(t')\ra$, can be obtained by making 
inaccurate measurements on a large ensemble of identical systems, without significantly perturbing individual
system's states.
Note, however, that according  to (\ref{h2}), the inverse problem has no solution for two-measurements one-step histories.
\section{Probabilities, averages, amplitudes}
There are three basic quantities in quantum mechanics that can be expressed in terms of each other, sometimes in a confusing manner.
To begin with, probabilities (frequencies) and probability amplitudes, which occur in quantum mechanics, are very different quantities.
The former appear to belong to the realm of the observable, the latter to the realm of theory.
A product of amplitudes is another amplitude, as shown in Eq.(\ref{1}), and so is a sum of amplitudes. For example, 
\begin{eqnarray}\label{i1}
\sum_{i_n}A(b^{L}_{i_L} \gets...\gets b^{n}_{i_n}\gets...\gets b^{1}_{i_1})=
A(b^{L}_{i_L} \gets...\gets b^{n+1}_{i_{n+1}} \gets b^{n-1}_{i_{n-1}}\gets ...\gets b^{1}_{i_1})
\end{eqnarray}
is an amplitude for passing through all, but the $n$-th, states.
\newline
Similarly, from (ii) and (iii), a product of probabilities is, as it should be, another probability.
A sum 
\begin{eqnarray}\label{i2}
\sum_{i_n}P(B^{L}_{i_L} \gets...\gets B^{n}_{i_n}\gets...\gets B^{1}_{i_1})=
P(B^{L}_{i_L} \gets...\gets B^{n+1}_{i_{n+1}} \gets B^{n-1}_{i_{n-1}}\gets...\gets B^{1}_{i_1}), 
\end{eqnarray}
is the probability of having $L-1$ values, assuming that the values of $\B_n$ are, in principle, known.
[If they cannot be known in principle, the correct expression for $P(B^{L}_{i_L} \gets...\gets B^{n+1}_{i_{n+1}} \gets B^{n-1}_{i_{n-1}} \gets ...\gets B^{1}_{i_1})$ would be $|A(b^{L}_{i_L} \gets...\gets b^{n+1}_{i_{n+1}} \gets b^{n-1}_{i_{n-1}}\gets ...\gets b^{1}_{i_1})|^2$, provided
all eigenvalues are distinct.] A real non-negative probability $P$ cannot, in general, equal a complex valued amplitude $A$, or even 
its real or imaginary part, since $\R [A]$ and $\text{Im}[A]$ can be both positive or negative. The only trustworthy connection between an $A$ and a $P$ is through the Born rule, which states that a probability must be an absolute square of an amplitude.
\newline
The third kind of quantities occurring in quantum mechanics are averages, or mean values.
For example, the mean value of an quantity $\B^n$, measured at $t=t_n$, would be
\begin{eqnarray}\label{i3}
\la \B^n\ra=\sum_{i_1, ...,i_L}B^n_{i_n}P(B^{L}_{i_L}\gets...\gets B^{1}_{i_1}).\n 
\end{eqnarray}
Since the eigenvalues $B^n$ can, in principle, have either sign, the only restriction on  $\la \B^n\ra$ is that 
it must be real valued. At first glance, the only information a $\la \B^n\ra$ can reveal relates to the absolute squares of the amplitudes,
contained in the $P(B^{L}_{i_L}...\gets B^{1}_{i_1})$'s .
This is, however, not the case for composite systems, because of the Superposition Principle (iv).
One example was given by Eq.(\ref{g2}), and we briefly review here its derivation. 
First, we considered two-step paths, $\(d_k,f' \gets c_j,f \gets b_i,G\)$, and asked for the value of the intermediate pointer position 
$f$,  regardless of its final position, given that at $t=t'$ the system reached $|d_k\ra$, passing first through $|c_j\ra$.
There are $N$ system+pointer states, $|f\ra|c_j\ra$,  which correspond to the same pointer position $f$. By  (iv), the 
probability of having first $f$, then $f'$, and the system in $|d_k\ra$, is found by adding the amplitudes for going through 
each of $|f\ra|c_j\ra$, taking absolute square of the result, and summing it over all $f'$. Thus, the values of $\C$, 
which belong to the system's past, remain indeterminate, except in the ``strong" limit  $G^2(f-C_j) \to \delta (f-Cj)$,
where interference is completely destroyed.
Information about the phases of $A^s(d_k\gets c_j\gets b_j)$ is passed to the mean pointer reading $\la f \ra$, 
whose relation with the amplitudes takes its simplest form, (\ref{g2}) and (\ref{g4}), in the ``weak" limit $\Delta f \to \infty$.
In this case, Eq.(\ref{g2}) relate a mean value in its l.h.s. to a real part of an amplitude in its r.h.s.
\newline
In a similar way, Eq.(\ref{h5}) equates a mean value to an amplitude, since $\la \psi|\C|\psi\ra$ is a scalar product 
of two states, $|\psi\ra$ and $\C|\psi\ra$. The mean reading is defined by Eq.(\ref{h4}), and its derived form in Eq.(\ref{h5})
is always valid, since for a Hermitian $\C$ the r.h.s. of (\ref{h5}) is real valued.
The amplitude $\la \psi|\C|\psi\ra$ can sometimes be used to evaluate a different probability.
\newline
For example, an experimentalist 
can measure $\C =\sigma_x$ for a qubit in some state $|\psi\ra=\cos \theta |\up _z\ra +\exp(i\varphi) \sin \theta |\do_z\ra$, to an arbitrary accuracy $\Delta f$. Having evaluated the average pointer reading,
\begin{eqnarray}\label{i4}
\la f\ra_{\C}=\la \psi |\sigma_x|\psi\ra,
\end{eqnarray}
he/she also knows the scalar product $\la \phi|\psi\ra$, where $|\phi\ra \equiv \sigma_x|\psi\ra= \exp(i\varphi) \sin \theta |\up_z\ra+\cos \theta |\do _z\ra$, and $\la \phi |\phi\ra=1$. 
With it, the statistics of measuring 
any operator which can be written as $\C'=|\phi\ra C'_1\la \phi|+|\phi_\perp\ra C'_2\la \phi_\perp|$,
$\la\phi_\perp|\phi\ra=0$, $\la\phi_\perp|\phi_\perp\ra=1$ on $|\psi\ra$, are also known (see Fig.7). In particular, the probability, $P_{\C'} (C'_1)$, of obtaining in an accurate 
measurement a result $C'_1$ is given by
\begin{eqnarray}\label{i4}
P_{\C'} (C'_1)= \la \phi|\psi\ra^2=\la f\ra_{\C}^2,
\end{eqnarray}
while the likelihood of obtaining $C'_2$ is $P_{\C'} (C'_2)=1-P_{\C'} (C'_1)$. Evaluating a sum $\sum_{j=1}^2 C'_jP(C'_j)=
\la \psi|\C'|\psi\ra$ yields a new amplitude, $\la \phi'|\psi\ra$, for $\phi'=\C'|\psi\ra$, and so on.
The scheme is of little practical use, since to identify the operator $\C$ one needs to know the state $|\psi\ra$ and, with it, 
the outcomes of all possible future measurements are known already. It does, however, 
illustrate 
the two ways relation between the probability amplitudes and probabilities
(see Fig.1).  
\begin{figure}[h]
\includegraphics[angle=0,height= 5cm]{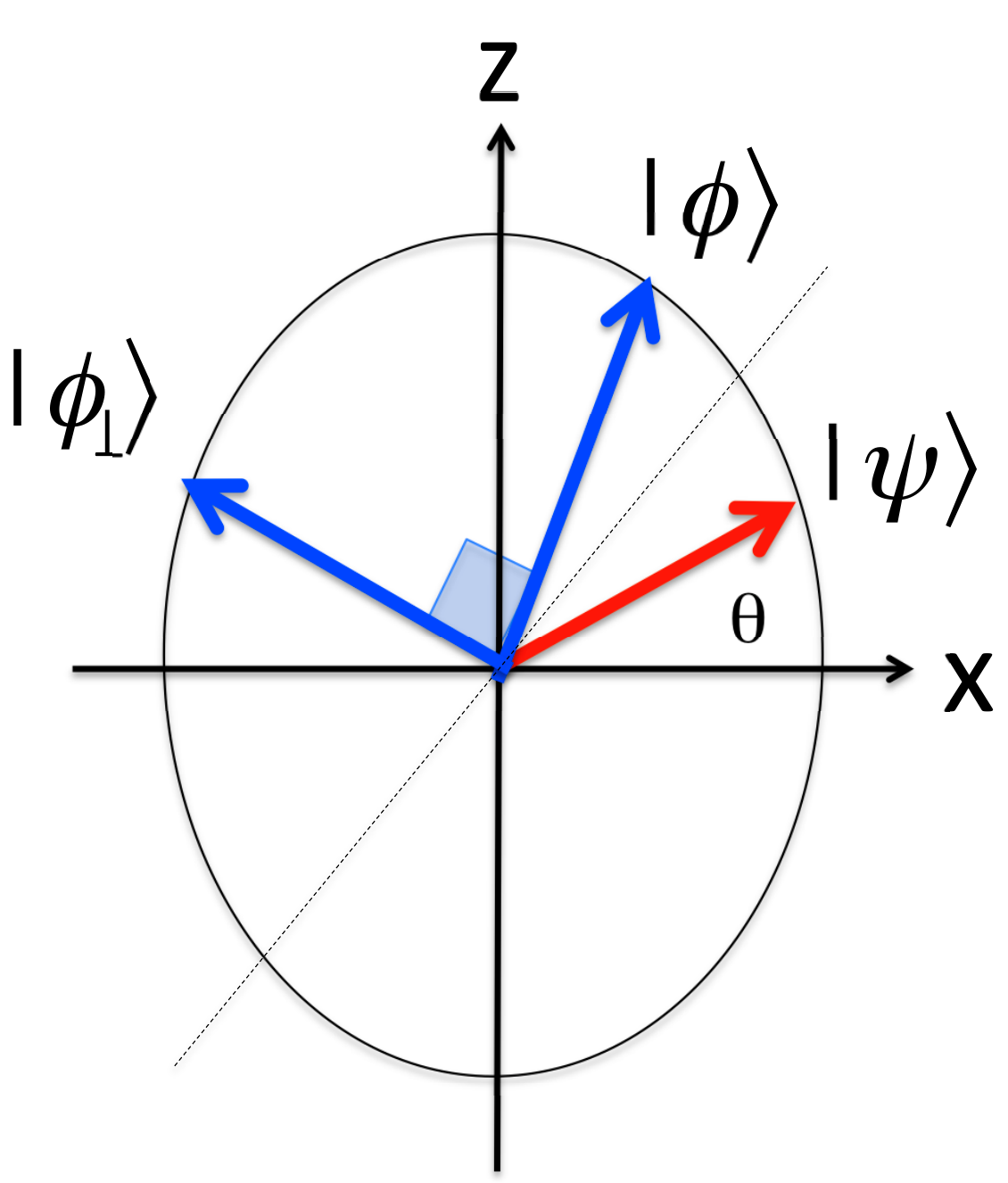}
\caption {A two-way relation between quantum theory and an experiment.
Calculated amplitudes predict the measured probabilities, while 
the measured probabilities allow one to evaluate some of the amplitudes involved.}
\label{fig:FIG1}
\end{figure}
\section{Amplitudes are amplitudes}
Suppose experimental data have been used, in one of the ways described above, to evaluate certain probability amplitudes. What exactly has been learnt? Clearly, the values of the amplitudes, and nothing new about their physical meaning, which remains as postulated in Section II. Quantum mechanics is a recipe for 
calculating frequencies of the outcomes which are observed, or can, in principle, be observed \cite{Feynl}.
The way to any observable frequency, inevitably passes through a complex valued amplitude, or amplitudes. 
Not surprisingly, sometimes it is possible to deduce the value of an amplitude from a measured frequency, 
although the task is complicated, by the Born rule, leading to the loss of information about certain phases.
\newline
The following example illustrates how the amplitudes should not be used.
Consider, first, a one-step case, where a three-state system, $N=3$, is prepared in a state $|b\ra$ at $t=0$, and  an operator, $\C$, with eigenvalues, $C_1=1$, $C_2=0$, while $C_3=0$, is accurately measured at $t=t'$ (see Fig.2a). Then the experiment is repeated with an operator 
$\C'$, which commutes with $\C$, and whose first two eigenvalues are the degenerate, $C'_1=1$, $C'_2=1$, and $C'_3=0$. 
Since in both cases  $t'$ is the final time, 
the probability to obtain $C'_1$ in the second 
case is just the sum of the probabilities to get $C_1$ and $C_2$ in the first measurement, 
\begin{eqnarray}\label{i4}
P(C'_1)=P(C_1)+P(C_2)=
|A^s(c_1\gets b)|^2+|A^s(c_2\gets b)|^2.
\end{eqnarray}
Renaming, for an additional effect,  the subspace spanned by $|c_1\ra$ and $|c_2\ra$ a "box", 
referring to each of the states as a "part of a box", and replacing the "system" with "particle", one concludes from Eq.(\ref{i4}) that
if a particle is in one part of the box, it is also in the box as a whole. 
This makes perfect classical sense. 
\newline
Next we add one more step, and consider an ensemble where the system is post-selected in 
a known final state $|d\ra$ at $t''>t'$. Now the time $t'$ belongs to the system's past, and the rules change.
In particular, for the two measurements we have
\begin{eqnarray}\label{i5}
P(C'_1)= 
|A^s(d\gets c_1\gets b) +A^s(d\gets c_2\gets b)|^2 \ne
 P(C_1)+P(C_2).
\end{eqnarray}
What is more, one is free to chose $A^s(d\gets c_1\gets b)=-A^s(d\gets c_2\gets b)\ne 0$, so that 
$P(C'_1)=0$.  Now the ``particle" can be found in one part of the ``box", but never in the ``box" as a whole.
\newline
There is, of course, no paradox. Different equipments, needed to measure $\C$ or $\C'$, 
affect the measured system in different ways, and create different ensembles \cite{DS3}.
This is obvious, since the probabilities of finding the system in $|d\ra$, are also different, 
\begin{eqnarray}\label{i6}
P(d)= \sum_{j=1}^3 |A^s(d\gets c_j\gets b)|^2\ne
P'(d)=|A^s(d\gets c_3\gets b)|^2.\q\n
\end{eqnarray}
 The conflicting statements refer to different circumstances, and should not be juxtaposed.  
\newline
The objection (\ref{i6}) can, however, be overcome, if we employ two inaccurate weak pointers 
to measure $\C$ and $\C'$. The pointers can be enacted simultaneously, and the probability to arrive
to $|d\ra$ can be affected as little as we want. As discussed in Section IX, the mean reading of the two pointers
are
\begin{eqnarray}\label{i7}
\la f\ra = \R \left [\frac{A^s(d\gets c_1\gets b)}{\sum_j A^s(d\gets c_j\gets b)} \right] \ne 0, \q \text{and} \q 
 \la f\ra = \R \left [\frac{A^s(d\gets c_1\gets b)+A^s(d\gets c_2\gets b)}{\sum_j A^s(d\gets c_j\gets b)} \right] = 0,
\end{eqnarray}
Reasoning as in \cite{Vaid}, one may say that the weak pointer ``has moved", if we look into the half of the box, 
but has not moved when we look into the whole box. Hence, there is a physical evidence that a quantum system
is in one part of the box, but not in the whole box, at the same time. 
This is, however, unfounded. We only confirmed the previously known relation between the amplitudes
$A^s(d\gets c_1\gets b)$ and $A^s(d\gets c_2\gets b)$, but have obtained no warranty \cite{DSphot} for using it as
an evidence of the ``presence" or ``absence" of the system in a particular "place"  (see Fig.8). 
While the values of the probability amplitudes can sometimes be deduced from the observable data, 
such an exercise gives no further clue as to why the amplitudes must play a fundamental in quantum mechanics.

\begin{figure}[h]
\includegraphics[angle=0, height= 5cm]{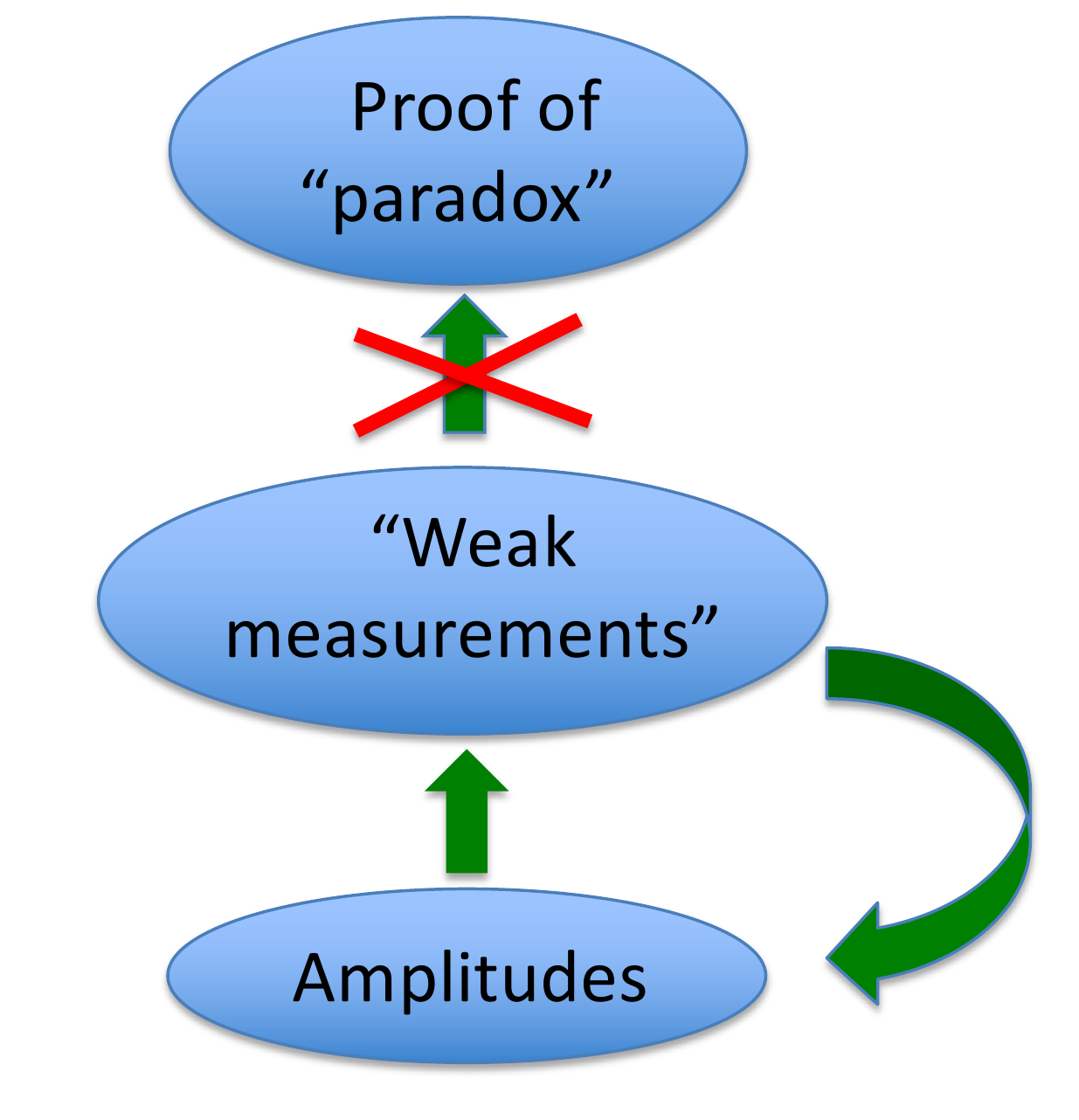}
\caption {Inaccurate ``weak measurements" do not prove that
conflicting properties of a system, 
subjected to different accurate measurements, 
are simultaneously present when the system is not being observed. 
Rather, they confirm the relations between the probability
amplitudes, often known to the experimentalist beforehand.
}
\end{figure}
\section{Summary and discussion}
Quantum theory predicts probabilities for the outcomes of series of consecutive instantaneous measurements 
made on elementary quantum systems. It does so by associating, with the measured quantities, 
 different bases in the Hilbert space, which represent the observed system.
Complex probability amplitudes, $A[path]$, are ascribed to
all virtual (Feynman) paths, connecting the basis states at the times of measurement. 
The amplitudes are then combined, according the degeneracies of the measured operators,
and the probability of a series of observed outcomes is obtained as the absolute square of the sum 
of the interfering amplitudes (Born rule). The shortest possible series is a two-measurement one-step sequence, 
in which the first measurement ``prepares" the system in a known initial state. 
The result is, therefore, a joint probability $P(C\gets B)$, to have an outcome $C$, given and earlier
outcome $B$. A failure to relate $B$ to a unique initial state $|b\ra$, would prevent the theory
from defining a statistical ensemble for the later outcome(s), thus depriving it of any predictive power.  
\newline
In this paper, our main purpose was to study the ways in which the information about the probability amplitudes 
can be obtained from the probability distributions, accessible to an experimentalist.
The Born rule seems to suggest that such information is irretrievably lost, in the act of taking the absolute value of an amplitude.
This is certainly true for a single degree of freedom. The difficulty can, however, be overcome for a system, $S_1$, which forms a part of a larger system, and is, therefore, coupled to something else, say $S_2$. This coupling needs to be of a particular form, so that the path amplitudes of the composite can be constructed with the help of the amplitudes of the measured system in isolation. If so, 
the role of the coupling is to affect the interference between the paths of $S_1$, and $S_2$ assumes the role of a measuring device. One, although not the only, example of such a device (see \cite{DSnh}), is the von Neumann pointer, described in Section IV. 
For a sequence involving more than two measurements (more than one step), accurate evaluation of the distribution of the pointer's readings allows one to reconstruct the path amplitudes of $S_1$, e.g.,  by one of the methods described in Sections V, VI, and IX.
One additional condition is that the resolution of the pointer should not be too high, or the phases of the amplitudes will be lost. 
In the case of pre- and post-selected system of Sections II-IX, the knowledge of all path amplitudes 
of the final state, and of he evolution operator, allows one to reconstruct also the unknown initial state \cite{{EXP2},{EXP3}}.
Other schemes, based on the response of a quantum system to a weak perturbation, can, in principle, be developed as tools 
for reconstructing different quantum probability amplitudes. 
\newline
We conclude with two general remarks. Firstly, even if probability amplitudes are considered to be a purely computational 
toll, it is hardly surprising that their values can be extracted from the observable probabilities. After all, one can arrive at the probabilities only through the amplitudes, so some sort of a reverse calculation is likely to be possible. 
Secondly, with our analysis, we are no closer to answering the more interesting question, namely why quantum theory requires the amplitudes in the first place? The need for a first observation, preparing the system in an initial state, before subsequent statistics can be predicted and may point towards limitations of our perception of the physical world  \cite{FOOT} . 
 \section{Appendix. Extension to mix states}
The analysis is readily extended to the case where the system is prepared in a mixed, rather 
than pure, state, 
\begin{eqnarray}\label{A1}
\hat R_i=\sum_\alpha w(\alpha) |b_i^\alpha\ra \la b_i^\alpha|,
\end{eqnarray}
by taking care of the normalisation factor $\N(d_k,b_i)$ in Eq.(\ref{d3}), since 
the new distribution of the meter readings, $\la \rho(f)\ra$ is not simply 
$\sum_\alpha \rho^\alpha(f)$. 
Below we consider the $L=3$ case, but the results are easily generalised for any number of measurements.
For the probability to find the pointer in $f$, and the 
system, prepared in $\hat R_i$, in $|d_k\ra$ we can, however, write
\begin{eqnarray}\label{A2}
\la \rho(f,d_k)\ra=\frac{\sum_\alpha w(\alpha)\N^\alpha (d_k,b_i)\rho^\alpha(f,d_k)}{\sum_\alpha w(\alpha)\N^\alpha (d_k,b_i) }.
\end{eqnarray}
Similar expressions are obtained for the moments of $\la \rho(f,d_k)\ra$, $\la f^n\ra\equiv \int df  f^n\la \rho(f,d_k)\ra$.
For example, in the accurate ``strong" limit, $\Delta f \to 0$ ,we have
\begin{eqnarray}\label{A3}
\la f \ra\to \frac{\sum_{j=1}^N C_j \left [\sum_\alpha  w(\alpha)|A^s (d_k \gets c_j\gets b^\alpha_i)|^2\right ]}
{\sum_{j=1}^N\sum_\alpha w(\alpha)|A^s (d_k \gets c_j\gets b^\alpha_i)|^2 }, \q
\end{eqnarray}
while inaccurate WM, $\Delta f \to \infty$, would yield
\begin{eqnarray}\label{A4}
\la f \ra\to \frac{\sum_{j=1}^N C_j \left [\sum_\alpha  w(\alpha) A^{s*} (d_k \gets b^\alpha_i)A^s (d_k \gets c_j\gets b^\alpha_i)\right ]}
{\sum_{j=1}^N\sum_\alpha w(\alpha)|A^s (d_k \gets b^\alpha_i)|^2 }, \q\q
\end{eqnarray}
where $A^s (d_k \gets b^\alpha_i)\equiv \sum_{j=1}^N A^s (d_k \gets c_j\gets b^\alpha_i)=\la d_k|\u^s(t'')|b^\alpha_i\ra$.
\begin{center}
\textbf{Acknowledgements}
\end{center}
Financial support  through the MCIU grant
PGC2018-101355-B-100(MCIU/AEI/FEDER,UE) and the Basque Government Grant No IT986-16.
is also acknowledged by DS.


\begin{thebibliography}{10}
\bibitem{Feynl} R.P, Feynman, R. Leighton, and M. Sands, {The Feynman Lectures on Physics III} (Dover Publications, Inc., New York, 1989), Ch.1: Quantum Behavior.
\bibitem{Feyn} R.P. Feynman and A.R. Hibbs,
{Quantum Mechanics and Path Integrals} (McGraw-Hill, New York 1965).
\bibitem{Leif} M.S.Leifer, 
Quanta \textbf{3}, 69, (2014).
\bibitem{WVrev} J. Dressel, M. Malik, F.M. Miatto, A.N. Jordan, \& R.W. Boyd, 
Rev. Mod. Phys. \textbf{86}, 307 (2014).
\bibitem{DS1} D. Sokolovski, 
Phys. Lett. A \textbf{380}, 1593 (2016).
\bibitem{DS2} D. Sokolovski, E. Akhmatskaya, 
Ann. Phys.\textbf{388}, 382 (2018).
\bibitem{WM1} Y. Aharonov, A. Botero, S. Popescu, B. Reznik, J. Tollaksen, 
Phys. Lett. A, \textbf{301}, 130 (2002).
\bibitem{WM2}  Y. Aharonov and L. Vaidman, in { Time in Quantum Mechanics}, edited by
J.G. Muga, R. Sala Mayato and I.L. Egusquiza (Second ed.,Springer, 2008), p. 399.
\bibitem{DS3} D. Sokolovski,  
Ann. Phys., \textbf{397}, 474 (2018).
\bibitem{EXP1}  J.S. Lundeen \& A.M. Steinberg, 
Phys. Rev. Lett.   \textbf{102}, 020404 (2009).
\bibitem{EXP2} J.S. Lundeen, B. Sutherland, A. Patel, C. Stewart,  \& C. Bamber, 
Nature \textbf{474}, 189 (2011). Note that, contrary to the claim made in the title, the authors measure not the wavefunction, but a relative amplitude similar to the one in Eq(\ref{g3}). In \cite{EXP2}, a mistake was made in going from Eq(6) to Eq(7). 
\bibitem{EXP3}  S. Kocsis, B. Braverman, S. Ravets, M.J. Stevens, R.P. Mirin, L. Krister Shalm, and A.M. Steinberg, 
Science \textbf{332}, 1170 (2011).
\bibitem{EXP4} M. Hallaji, A. Feizpour, G. Dmochowski, J. Sinclair, and A.M. Steinberg,
Nature Physiscs   \textbf{13}, 550 (2017).
\bibitem{MW} L. Vaidman, { Many-Worlds Interpretation of Quantum Mechanics}, The Stanford Encyclopedia of Philosophy (Fall 2018 Edition), Edward N. Zalta (ed.), URL = https://plato.stanford.edu/archives/fall2018/entries/qm-manyworlds/
\bibitem{vN} J. von Neumann, {\it Mathematical Foundations of Quantum Mechanics} (Princeton University Press, Princeton, 1955), pp. 183-217.
\bibitem{DSnh} D. Sokolovski, 
Phys. Rev. A  \textbf{66}, 032107 (2002).
\bibitem{CLT} J. Rice, {\it Mathematical Statistics and Data Analysis}. 3rd edition, (Duxbury Advanced, 2010).
\bibitem{AV} Y. Aharonov, D.Z. Albert,  \& L. Vaidman, 
Phys. Rev. Lett. \textbf{60}, 1351 (1988).
\bibitem{Merm} N.D. Mermin, 
Rev. Mod. Phys., \textbf{65}, 803 (1993).
\bibitem{3box} See, for example, T. Ravon, L. Vaidman, 
J. Phys. A: Math. Theor., \textbf{40}, 2873 (2007).
\bibitem{Vaid}  L. Vaidman, 
Phys. Rev. A, {\bf 87}, 052104 (2013).
\bibitem{DSphot} D. Sokolovski, 
Phys. Lett. A, {\bf 381}, 227, (2017).
\bibitem{FOOT} Suppose the proverbial Alice receives a single spin-1/2 in an unknown state. The spin performs Rabi oscillations in a magnetic field,  directed along the $z$-axis,
and she wants to know the odds on finding it directed up the $x$-axis after a given time. The no-cloning theorem prohibits determining the spin's state, and Alice has no other choice but to measure, say,  the spin's $x$-component. It is only then that she can be certain that, after 
the Rabi period, the spin will be found pointing in the same direction.  However, it has been shown \cite{Merm}, that the initial value
of the spin's projection cannot pre-exist the first Alice's measurement, and the unknown state the spin was supposed to be in initially, 
was probably destroyed by it. We will never know, and might as well drop the very concept of an unknown state no one has ever seen.
The state ascribed to the spin after Alice has obtained her first result is, on the other hand, useful, since now the statistics for all measurements, which are to follow, can be obtained. But the very idea of a quantum state appears to be related to what Alice {\it perceives},  rather that to what the spin really {\it is}.
\end{thebibliography}
\end{document}